\documentclass[twocolumn,showpacs,preprintnumbers,amsmath,amssymb]{revtex4-1}

\usepackage{graphicx}	% Include figure files
\usepackage{dcolumn}		% Align table columns on decimal point
\usepackage{bm}			% bold math
\usepackage[dvips]{epsfig}
\usepackage[latin1]{inputenc}
\usepackage{color}

\newcommand \be {\begin{equation}}
\newcommand \ee {\end{equation}}
\newcommand \bea {\begin{eqnarray}}
\newcommand \eea {\end{eqnarray}}

\begin{document}

\title{Aging of the frictional properties induced by temperature variations}

\author{Jean-Christophe G\'eminard and Eric Bertin}
\affiliation{Universit\'e de Lyon, Laboratoire de Physique, Ecole Normale Sup\'erieure de
Lyon, CNRS, 46 All\'ee d'Italie, 69364 Lyon cedex 07, France.}

\begin{abstract}
The dry frictional contact between two solid surfaces is well-known to obey Coulomb friction laws.
In particular, the static friction force resisting the relative lateral (tangential) motion of solid
surfaces, initially at rest, is known to be proportional to the normal force and independent of the
area of the macroscopic surfaces in contact.
Experimentally, the static friction force has been observed to slightly depend on time.
Such an aging phenomenon has been accounted for either by the creep of the material
or by the condensation of water bridges at the microscopic contacts points.
Studying a toy-model, we show that the small uncontrolled temperature changes of the system
can also lead to a significant increase of the static friction force. 
\end{abstract}

\maketitle

\section{Introduction.} 

Granular materials, or more generally macroscopic solids in frictional contact,
at mechanical equilibrium are {\it a priori} considered as athermal systems,
meaning that thermal agitation has no significant feedback on the mechanical degrees
of freedom at the grain scale. 
However, uncontrolled temperature variations inevitably produce dilations of materials
and one can wonder whether they can alter the properties, in particular the mechanical
stability, of such systems.

Several recent studies showed that temperature cycles, even of small amplitude, can
induce the slow compaction of dry granular materials \cite{Chen06,Chen09,Divoux08,Divoux09}.
The mechanisms, in particular the role played by the confining walls, are still under debate
but one can anyway conclude that the temperature variations cause {\it aging}: 
the properties of the material, here at least the density, evolve in an irreversible manner with time.

In granular matter, several sources of aging were identified.
First, for \lq\lq dry\rq\rq granular materials in a humid atmosphere,
the condensation of microscopic liquid bridges at the contacts between grains,
leads to a cohesive force \cite{Bocquet98}:
the nucleation of the bridges being an activated process, the mechanism leads to a logarithmic increase of the angle of avalanche \cite{Bocquet98,Fraisse99} and of the static frictional coefficient \cite{Losert00,Geminard01} with time. Second, for immersed granular materials, chemical reactions at the surface of the grains results in soldering them to each other \cite{Fraisse99}. Again, the angle of avalanche \cite{Gayvallet02} and the static frictional coefficient \cite{Losert00,Geminard01} are observed to increase logarithmically with time. 

In the solid friction \cite{Bowden50,Rabinowicz65}, which characterizes the mechanical contact between solid surfaces, a similar aging phenomenon also takes place: the static frictional coefficient is also observed to increase logarithmically with time for contacts between various materials like metals \cite{Rabinowicz58}, rocks \cite{Scholz90}, Bristol paper \cite{Heslot94}. Again, for \lq\lq dry\rq\rq  friction in a humid atmosphere, the condensation of liquid bridges at the micro-contacts between the flat, but nevertheless rough surfaces, can account for a part of the phenomenon \cite{Crassous99}. However, when effects of the humidity are suppressed, aging is still observed. The phenomenon is due to the creep of the material at the micro-contacts, as proven by the dependence of the aging dynamics on the overall temperature of the system \cite{Berthoud99}. Experiments even exhibited a logarithmic aging of the micro-contacts themselves \cite{Bureau02}. 

To our knowledge, the possibility for the temperature variations to be responsible for a part of the aging observed in solid friction was still not evaluated.
We can guess that the dilation or contraction of the materials which result from a change in the temperature alter the stress distribution between the micro-contacts
and that, as a result, the system can evolve with time due to a pinning-depinning dynamics. Such dynamics of the micro-contacts has been proposed to explain the
dissipation associated with the relative motion of the surfaces and, thus, to account for the dynamic friction \cite{Tomlinson29,Joanny84,Caroli96} but effects of
dilations have not been evaluated. The study of the hysteresis cycle of the tangential force induced by a quasi-static cyclic displacement
is a related but slightly different problem \cite{Crassous97}: small displacements, even small enough not to be considered as macroscopic sliding,
induce irreversible micro-slips and a slow evolution of the system with time \cite{Olofsson95}.
The temperature variations would have a slightly different effect as they rather induce random stress variations, correlated at long range.

{In the present article, we propose the study of a minimal model mimicking the frictional contact between two, nominally flat, solid surfaces, a slider on top of a horizontal substrate.
We first consider that, due to roughness, the real contact between the surfaces reduces to a large, but finite, number of microcontacts,
themselves belonging to a small number of mesoscopic, coherent, contact regions \cite{Persson10}.
Such coherent regions are composed of a large number of microcontacts, so that their contact with the substrate obeys the Amonton-Coulomb
law for static friction. However, the number of microcontacts in a coherent region is finite \cite{Baumberger06} so that the associated static friction coefficient,
$\mu_\mathrm{s}$, is sensitive to the local microstructures in regard and, thus, depends on the position on the substrate.
In addition, coherent regions are elastically connected one to another by the material the slider and the substrate are made of.}

{In order to mimic the practical situation in the simplest model,  we further reduce the problem to the study of a system
consisting of two sliders connected by a linear spring. Each of the sliders accounts for one of the coherent regions in frictional contact with a flat substrate
whereas the spring accounts for the elastic coupling between them.
We shall demonstrate that, when the system is subjected to thermal dilations, the mechanical stability of the slider, accounted for by an effective
static frictional coefficient, exhibits a significant increase with time due to the distributed values of the static friction coefficient $\mu_\mathrm{s}$.}

\section{The model}

\subsection{Description}

\begin{figure}[!h]
\includegraphics[width=\columnwidth]{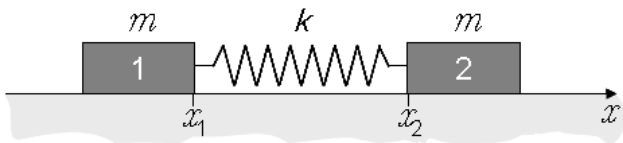}
\caption{\label{sketch}{Sketch of the considered situation.}}
\end{figure}

The system (mass $M$), subjected to its own weight $Mg$, lies on a flat and horizontal surface. In a simple approach,
we further assume that each contact region sustains half the weight of the system, so that the normal force acting on one slider is $N = Mg/2$. We introduce the rest distance $l_0$ between the two contact regions at the temperature $T_0$. Due to the elasticity of the slider material (assumed to be much softer than the substrate), the small sliders are connected by a spring of length $l_0$ and stiffness $k$ and, being identical, they are associated with the same inertia, {\it i.e.} the same mass $m \equiv M/2$. With these assumptions, the system reduces to two sliders, connected by a spring, moving on a flat surface as sketched in the figure \ref{sketch}. We denote by $x_1$ and $x_2$ the positions of the small sliders 1 and 2 respectively such that the length of the spring $l = x_2 - x_1$.

At rest, the contact between {each slider} and the substrate is characterized by the static frictional coefficient $\mu_\mathrm{s}$,
{which accounts for the value that the horizontal force must overcome to move the slider.}
Due to the {heterogeneity} of the surfaces in regard, the local static frictional coefficient might take, at random, different values, $\mu_{\mathrm{s},1}$ and $\mu_{\mathrm{s},2}$, for the two contact regions.
By contrast,
{we characterize the dynamical friction by a single constant value of $\mu_\mathrm{d}$. Indeed, the surface explored by the sliders being homogeneous
in average, the dissipative interaction with the substrate can reasonably be characterized,
over the whole surface, by a single dynamical frictional coefficient, $\mu_\mathrm{d}$, which quantifies the rate of energy dissipation.
We further assume, in agreement with standard observations, that $\mu_\mathrm{d} < \mu_\mathrm{s}$.}

The dynamics of the system is induced by temperature changes which, due the thermal dilation of the materials, lead to changes in the natural length of the spring $l$ according to:
\begin{equation}
l(T) = l_0 \bigl[ 1 + \kappa (T - T_0) \bigr]
\end{equation}
where $T$ stands for the temperature and $\kappa$ for the thermal expansion coefficient of the slider material. We assume, for the sake of simplicity, that the substrate does not dilate, so that the positions $x_1$ and $x_2$ are not altered by the temperature changes, and that the spring constant does not depend on the temperature. Thus, due to the thermal dilation, the spring expands or contracts so that the slider 2 is subjected to the tangential force 
\begin{equation} \label{eqF12}
F_{1\to2} = - k ( x_2 - x_1 - l ) 
\end{equation}
where we assume the $x$-axis to be oriented from the slider 1 to the slider 2. In the same way, the slider 1 is subjected to $F_{2\to1}=-F_{1\to2}$. If $|F_{1\to2}|$ exceeds one of the static friction forces, $\mu_{\mathrm{s},1} N$ or $\mu_{\mathrm{s},2} N$, the corresponding slider starts moving and the system rearranges. After the slider has stopped, the corresponding value of the static frictional coefficient has changed because the slider lies at a different position on the substrate. The new value of the frictional coefficient is drawn at random from a distribution $p(\mu_\mathrm{s})$.
In addition, we assume for consistency that $\mu_\mathrm{s}$ is always larger than the dynamic frictional coefficient $\mu_\mathrm{d}$, namely $p(\mu_\mathrm{s})=0$ for $\mu_\mathrm{s} < \mu_\mathrm{d}$.

The aim of this study is to account for the evolution of the system properties, thus of $\mu_{\mathrm{s},1}$ and $\mu_{\mathrm{s},2}$, as time elapses (aging).  

\subsection{System of equations}

The problem reduces to the study of two sliders, connected by a spring, in frictional contact with a substrate.

First, one can write the equations governing the dynamics of the sliders when in motion.
Introducing the thermal dilation $\theta\equiv \kappa (T - T_0)$, one gets
\begin{eqnarray}
m\ddot{x}_1 = - k[ x_1 - x_2 + l_0(1 + \theta)] - \mu_\mathrm{d} N S(\dot{x}_1)
\label{dynamics}\\
m\ddot{x}_2 = - k[ x_2 - x_1 - l_0(1 + \theta)] - \mu_\mathrm{d} N S(\dot{x}_2)
\nonumber 
\end{eqnarray}
where $S$ denotes the sign function.
The dynamics is characterized by a typical time scale
$\tau_\mathrm{dyn} = \sqrt{m/k}$.
We remark that, in practice, $\tau_\mathrm{dyn} \sim l_0/c_s$
where $c_s$ stands of the speed of sound in the material the macroscopic solid is made of.
For a typical size $l_0 \sim 10$~cm and usual values of $c_s$ (about a few kilometers per second),
we estimate $\tau_\mathrm{dyn} \sim 10^{-6} - 10^{-5}$~s.

Second, the static solid-friction is accounted for by the coefficients $\mu_{\mathrm{s},j}$ ($j=1,2$) such that, initially at rest, the slider $j$ starts moving and, thus, the system rearranges when
\begin{equation}
k \bigl| x_2 - x_1 - l_0(1+\theta) \bigr| = \mu_{\mathrm{s},j} N.
\label{stability}
\end{equation}

\subsection{Rearrangements}

From now on, we assume that after $n$ rearrangements, the sliders are at the positions
$x_1^n$ and $x_2^n$ associated with the values $\mu_{\mathrm{s},1}^n$ and $\mu_{\mathrm{s},2}^n$ of the static frictional coefficients. From the condition \eqref{stability}, one notes that the rearrangements involve the weakest slider, namely the one associated to the smallest value of the static frictional coefficient. For convenience, we denote: 
\begin{equation}
\mu_\mathrm{min}^n \equiv \min(\mu_{\mathrm{s},1}^n,\mu_{\mathrm{s},2}^n)
\equiv \mu_{\mathrm{s},i_n}^n
\end{equation}
where we introduce the index $i_n$ of the corresponding slider. With these definitions, we get the value $\theta_\mathrm{c}^n$ of the dilation $\theta$ at which the $(n+1)^\mathrm{th}$ rearrangement occurs:
\begin{equation}
\theta_c^n = \frac{x_2^n - x_1^n}{l_0} - 1 + \mu^n_\mathrm{min}\, \tilde{N}S(\dot\theta)
\label{critique}
\end{equation}
where the dimensionless normal force $\tilde{N}=N/(kl_0)$ has been introduced
to lighten the notations.
Eq.~(\ref{critique}) holds true when the system dilates ($\dot\theta > 0$) or contracts ($\dot\theta < 0$), the sign $S(\dot\theta)$ being evaluated before the rearrangement, for $\theta \to \theta_c^n$.

For $\theta = \theta_c^n$, the slider $i_n$ moves and reaches a novel static position such that:
\begin{equation}
\Delta x_{i_n} = 2 (-1)^{i_n} ( \mu^n_\mathrm{min} - \mu_\mathrm{d} ) \tilde{N} l_0 S(\dot\theta).
\label{displacement}
\end{equation}
One can easily show that the second slider necessarily remains at rest so that the rearrangement induces a displacement:
\begin{equation}
\Delta x_G^{n} = (-1)^{i_n} ( \mu^n_\mathrm{min} - \mu_\mathrm{d} ) \tilde{N} l_0 S(\dot\theta)
\label{displacmt-G}
\end{equation}
of the center of mass $G$ of the two sliders.

In conclusion, when the dilation $\theta$ reaches the critical value $\theta^n_c$, the weakest slider $i_n$ moves by $\Delta x_{i_n}$ and reaches a novel static position whereas the other slider remains at rest. As a consequence, the static frictional coefficient $\mu_{\mathrm{s},i_n}$ takes a new random value, $\mu_{\mathrm{s},i_n}^{n+1}$, associated with the new position $x_{i_n}^{n+1} = x_{i_n}^{n} + \Delta x_{i_n}$. The position of the other slider and the corresponding static frictional coefficient  remain unchanged.

\section{Numerical analysis}
\label{numerical}

We shall consider the temporal evolution of the system when subjected to aleatory temperature changes.
In the section \ref{changes}, we describe how the temperature changes are accounted for and, then, in the section \ref{results} we report the behavior of the system.
For practical purposes, we consider in this section a Gaussian distribution
$p(\mu_\mathrm{s})$, namely
\begin{equation}
p(\mu_\mathrm{s}) = \frac{1}{\sqrt{2\pi\sigma_\mu^2}} \exp\left[-\frac{( \mu_\mathrm{s} - \overline{\mu}_\mathrm{s})^2}{2 \sigma_\mu^2}\right].
\end{equation}
Although this distribution does not strictly speaking satisfy the condition
$p(\mu_\mathrm{s})=0$ for $\mu_\mathrm{s}<\mu_\mathrm{d}$, this condition will be
satisfied in practice as we restrict our study to the case
$\sigma_\mu \ll (\overline{\mu}_\mathrm{s}-\mu_\mathrm{d})$.

\subsection{Temperature changes}
\label{changes}

We shall assume that the temperature $T$ of the system fluctuates around the temperature $T_0$ such that $\theta$ is distributed according to a Gaussian distribution
$\psi(\theta)$ with zero mean and variance $\sigma_{\theta}^2$.
In addition, due to its thermal inertia, the system exhibits a thermal characteristic time $\tau_\mathrm{th}$ which limits the dynamics of the temperature changes and, thus, of the dilations. We assume that the temperature variations occur on time scales much
larger than the typical time scale of the rearrangement dynamics, that is,
$\tau_\mathrm{th} \gg \tau_\mathrm{dyn}$. 

In order to account for the finite value of $\tau_\mathrm{th}$, we consider the evolution of the system in the following way: At time $t_q = q\,\tau_\mathrm{th}$ ($q$ integer), the system is described by the positions $x_1^n$ and $x_2^n$, the frictional coefficients $\mu_{\mathrm{s},1}^n$ and $\mu_{\mathrm{s},2}^n$ and the dilation $\theta(t_q)$. 
A new value of $\theta(t_{q+1})$ is drawn at random from the distribution $\psi(\theta)$.
Then, we consider $\dot\theta$ as a constant over the time interval
$[t_q,t_{q+1}]$, namely
\be
\dot\theta = \frac{\theta(t_{q+1}) - \theta(t_{q})}{\tau_\mathrm{th}}
\ee
and one can calculate $\theta_c^n$ from Eq.~(\ref{critique}). The corresponding positions $x_1^{n+1}$ and $x_2^{n+1}$ after the rearrangement are calculated according to Eq.~(\ref{displacement}), and the smallest frictional coefficient among $\mu_{\mathrm{s},1}^{n}$ and $\mu_{\mathrm{s},2}^{n}$ are drawn anew from $p(\mu_s)$, the other remaining unchanged. The process is repeated while $\theta_c^n \le \theta(t_{q+1})$ if $\dot\theta>0$, or $\theta_c^n \ge \theta(t_{q+1})$ if $\dot\theta<0$.
When the evolution stops, the dilation between $t_q$ and $t_{q+1}$ has led to $r$ rearrangements. The system has reached a new state and the process can be iterated,
by choosing randomly a new value $\theta(t_{q+2})$.

\subsection{Numerical results}
\label{results}

\begin{figure}[!h]
\includegraphics[width=\columnwidth]{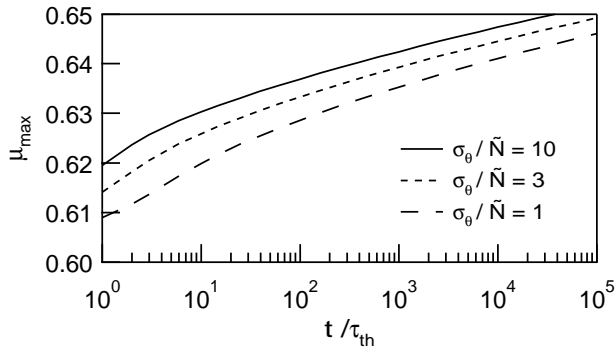}
\caption{\label{aging}{Maximum frictional coefficient $\mu_\mathrm{max}$ vs.~dimensionless time $t/\tau_\mathrm{th}$} - The maximum $\mu_\mathrm{max}$ is observed to increase almost logarithmically with time $t$, the aging being more effective when the amplitude $\sigma_\mathrm{\theta}$ of the temperature variations is larger ($\mu_\mathrm{d} = 0.5$, $\overline{\mu}_\mathrm{s} = 0.6$, $\sigma_\mathrm{\mu} = 0.01$, $\tilde{N} = 10^{-7}$).}
\end{figure}

When the system is subjected to temperature changes, the values of the static frictional coefficients continuously change in time.
As the rearrangements only involve the weakest slider, the maximum of the values $\mu_\mathrm{s,1}^n$ and $\mu_\mathrm{s,2}^n$,
\begin{equation}
\mu_\mathrm{max}^n \equiv \max(\mu_{\mathrm{s},1}^n,\mu_{\mathrm{s},2}^n),
\label{mumax}
\end{equation}
cannot decrease, and thus on average continuously increases with time $t$ (Fig.~\ref{aging}). 
We indeed observe that $\mu_\mathrm{max}$ increases rather logarithmically with time $t$ 
and that, for instance, the {\it aging} is faster for larger $\sigma_\theta$. 

\begin{figure}[!h]
\includegraphics[width=\columnwidth]{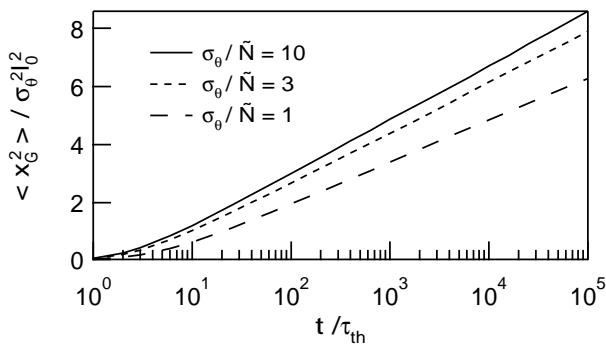}
\caption{\label{diffusion}{Normalized mean square displacement $\langle x_\mathrm{G}^2 \rangle / \sigma_\mathrm{\theta}^2 l_0^2$ vs.~dimensionless time $t/\tau_\mathrm{th}$} - The mean square displacement $\langle x_\mathrm{G}^2 \rangle$ increases logarithmically with time $t$, the diffusion being faster when the amplitude $\sigma_\mathrm{\theta}$ of the temperature variations is larger ($\mu_\mathrm{d} = 0.5$, $\overline{\mu}_\mathrm{s} = 0.6$, $\sigma_\mathrm{\mu} = 0.01$, $\tilde{N} = 10^{-7}$).}
\end{figure}

In addition, due to the random motion of the center of mass associated with the rearrangements (Eq.~\ref{displacement}), the system diffuses.
We report the mean square displacement $\langle x_\mathrm{G}^2 \rangle$ as function of time $t$ (Fig.~\ref{diffusion}) and observe that
$\langle x_\mathrm{G}^2 \rangle \sim \log(t/\tau_\mathrm{th})$, the diffusion being faster for larger $\sigma_\theta$.

In the next section, we shall discuss further these results theoretically,
describing analytically how the aging and the diffusion depend on the parameters of the problem
at long times.

\section{Theoretical analysis}
\label{theory}

\subsection{Dilation and rearrangements}

\begin{figure}[!h]
\includegraphics[width=\columnwidth]{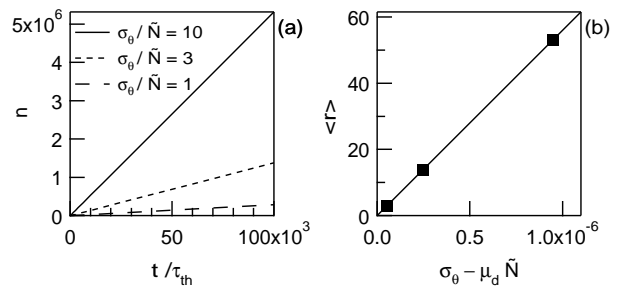}
\caption{{(a) Number of rearrangements $n$ vs.~time $t$ - The number of rearrangements $n$ increases, to within small fluctuations, linearly with the time $t$. (b) Average number of rearrangements $<r>$ during $\tau_{th}$ vs.~amplitude $\sigma_\mathrm{\theta}$ - In average, the slope $dn/dt$ or equivalently $<r>$ depends linearly on the amplitude $\sigma_\mathrm{\theta}$ according to Eq.~(\ref{slope}).}}
\label{rearrangement}
\end{figure}

In a first step, we consider the number $r$ of rearrangements that occur during the time interval $\tau_\mathrm{th}$. Considering the Eqs.~(\ref{critique}) and (\ref{displacement}), the difference between two successive values of the dilation at which rearrangements occur can be estimated as
\be
|\theta_c^{n+1} - \theta_c^{n}| \approx 2 (\overline{\mu}_\mathrm{s} - \mu_\mathrm{d}) \tilde{N} S(\dot\theta).
\ee
Thus, a dilation or a contraction of amplitude $\Delta\theta$ leads to a typical number of rearrangements $r \approx |\Delta\theta| / |\theta_c^{n+1} - \theta_c^{n} |$. 
To average over the fluctuations of $\Delta\theta$, we define it quantitatively
as the difference between two values $\theta$ and $\theta'$ randomly drawn
from $\psi(\theta)$. Hence $\Delta\theta$ is a Gaussian random variable with
zero mean and variance $2\sigma_{\theta}^2$.
We thus find for the average value $\langle r \rangle$
\be
\langle r \rangle \approx \frac{\sigma_\theta}{\sqrt{\pi} (\overline{\mu}_\mathrm{s} - \mu_\mathrm{d}) \tilde{N}}\, .
\ee
Comparison with the results of numerical simulations (Fig.~\ref{rearrangement})
essentially confirms this prediction. A careful analysis of the numerical data
however reveals the presence of a small shift. The average number of rearrangements
is found to be described by the phenomenological form
\begin{equation}
\langle r \rangle \approx \frac{\sigma_\theta - \mu_\mathrm{d} \tilde{N}}{\sqrt{\pi} (\overline{\mu}_\mathrm{s} - \mu_\mathrm{d}) \tilde{N}}\,.
\label{slope}
\end{equation}
The small offset $\mu_\mathrm{d} \tilde{N}$ is intuitively expected to result from 
the absence of rearrangements when $\dot\theta$ changes sign.

From now on, we shall consider the evolution of the system in terms of the number of rearrangements $n$. The temporal evolution of the system can then be assessed by considering the simple relation between $n$ and time $t$.

\subsection{Aging of the static frictional coefficient} 

It is particularly interesting to focus on the dynamics of $\mu_\mathrm{max}^n$
as this quantity plays an important role in the stability of the system (Sec.~\ref{discussion}).
Actually, one simply needs to notice that the dynamics consists in drawing randomly
two values $\mu_{\mathrm{s},1}^1$ and $\mu_{\mathrm{s},2}^2$ in the initial step $n=1$,
and then drawing at each step a new value of the frictional coefficient to replace
the smallest one. As a results, the value $\mu_\mathrm{max}^n$ is the maximum value
of the set of $n+1$ random values drawn independently.
This mechanism is actually very close to the one appearing in a standard
model of glassy dynamics, namely the Barrat-M\'ezard model
\cite{BarratMezard,Bertin10}.
The statistics of $\mu_\mathrm{max}^n$ then boils down to an extreme value problem
of independent and identically distributed random variables, for which standard results
are well-known \cite{Gumbel,Galambos}.
In particular, the distribution of $\mu_\mathrm{max}^n$ belongs for large $n$ to one of
the three classes of extreme value statistics, namely the Gumbel, Weibull or Fr\'echet class.
However, we are here more interested in the typical value of $\mu_\mathrm{max}^n$ rather
than by its relative fluctuations.
This typical value can be estimated through the following simple scaling argument.
We denote as $F(\mu_\mathrm{s})$ the complementary cumulative probability distribution
\be
F(\mu_\mathrm{s}) = \int_{\mu_\mathrm{s}}^{\infty} p(\mu) d\mu.
\ee
Having $\mu_\mathrm{max}^n$ smaller than a given value $\mu$ is equivalent to the fact that
the $n+1$ random values drawn dynamically are below the value $\mu$.
The corresponding probability is simply $[1-F(\mu)]^{n+1}$.
The typical value $\mu_\mathrm{typ}$ of $\mu_\mathrm{max}^n$ then satisfies
\be
[1-F(\mu_\mathrm{typ})]^{n+1} \approx \frac{1}{e}
\ee
which for large $n$ takes the form
\be
1-F(\mu_\mathrm{typ}) \approx e^{-\frac{1}{n+1}} \approx 1-\frac{1}{n}
\ee
so that finally $F(\mu_\mathrm{typ}) \approx 1/n$.
If the upper wing of the distribution $p(\mu)$ decays as $p(\mu) \sim e^{-c\mu^{\alpha}}$
where $c$ and $\alpha$ are positive constants
(for instance, $\alpha=2$ for a Gaussian distribution), then $F(\mu)$ behaves in the same way
as $p(\mu)$ up to algebraic prefactors. It follows that
\be
\mu_\mathrm{typ} \approx \overline{\mu}_\mathrm{s} + \sigma_\mu c^{-1/\alpha} (\ln n)^{1/\alpha}
\ee
to leading order in $n$.
Thus, in the case of the Gaussian distribution considered in
the section~\ref{numerical}, the simple theoretical argument presented here
predicts an increase of $\mu_\mathrm{max}$ like
\be
\mu_\mathrm{max} \approx \overline{\mu}_\mathrm{s} + \sigma_\mu \sqrt{2\ln
\left(\frac{\langle r\rangle t}{\tau_\mathrm{th}}\right)},
\label{mumaxth}
\ee
taking into account the correspondence $n \approx \langle r\rangle t/\tau_\mathrm{th}$
between the time $t$ and the total number $n$ of rearrangements (Eq.~\ref{slope}).

\subsection{Slowing down diffusion} 

In Sec.~\ref{numerical}, we obtained a logarithmic subdiffusion of the center of mass,
which we now would like to interpret.
We first observe that the evolution of the system can be divided in successive periods
during which the strongest slider remains the same, that is, the frictional
coefficient of the weakest slider is repeatedly drawn anew, without exceeding the
value $\mu_\mathrm{max}$ of the strong slider.
During such periods, the position of the weakest slider changes, due to the rearrangements,
and the net displacement $\ell$ is limited by the extension of the spring.
One can estimate $\ell \sim \sigma_\theta l_0$. 
%Hence, as a first approximation, the motion of the center of mass can be neglected
%in these periods.
When the chosen frictional coefficient becomes larger than the former
value $\mu_\mathrm{max}$ (which can be called a record breaking),
the weakest and strongest sliders get exchanged.
Hence the mean square displacement of the center of mass after $n$ rearrangements
should be of the order of $\ell^2$
times the number of record breakings in this sequence of rearrangements.

It is a classical result from statistical record theory \cite{Krug}
that the mean number of records that occur when successively
drawing $n$ independent and identically distributed random values
is, to leading order, equal to $\ln n$ for large $n$.
Further using the relation $n \approx \langle r\rangle t/\tau_\mathrm{th}$,
one eventually finds for the mean square displacement of the center of mass
\be
\langle x_G^2 \rangle \approx \sigma_\theta^2\,l_0^2  \ln\left(\frac{\langle r\rangle t}{\tau_\mathrm{th}}\right),
\ee
thus accounting for the logarithmic time dependence and the order of magnitude of the numerical data reported in Fig.~\ref{diffusion}.

\section{Discussion}
\label{discussion}

We have shown that the dilations associated with the temperature changes are likely to induce
the aging of the physical properties of the system (the maximum frictional coefficient increases with time and the diffusion slows down).
At this point, it is particularly interesting to discuss, first, the orders of magnitude of the temperature changes necessary for the
aging process to be at stake and, second, the possibility to assess the aging process experimentally.

Let us now consider the order of magnitude of the dilations likely to induce the aging. 
From equation (\ref{slope}), given that $\mu_\mathrm{d} \sim 1$, we deduce that the minimal amplitude of the dilations necessary to rearrange the system
is of the order of $\tilde{N}$  which can be estimated as follows.
Consider that the two contact points, sketched by the slider 1 and 2, are located at the distance $l_0$ one from the another.
Taking into account the Young modulus, $Y$, the material is made of, one can estimate $k \simeq Y A / l_0$ where $A$ stands
for the surface area of the slider cross section in the perpendicular direction. Considering the density $\rho$ of the material and
volume $V \sim A l_0$ of the slider, one can estimate the normal force, due to gravity $g$, $N \sim \rho A l_0\,g$.
Thus, we get $\tilde{N} \sim \rho g l_0 / Y$.
In practice, considering $g = 10$~m.s$^{-2}$, $\rho \sim 10^3$~kg.m$^{-3}$, $Y \sim 100$~GPa and the typical size $l_0 = 10$~cm, we get $\tilde{N} \sim 10^{-8}$. Thus, dilations of the order of $10^{-8}$ are likely to rearrange the system. Such dilations, the typical thermal expansion coefficients being of about $10^{-5}$~K$^{-1}$, correspond to temperature changes of about 10$^{-3}$ K. Thus, in usual experimental conditions, the thermal dilations produce rearrangements at the contact scale. Even more, it is not obvious to perform a control of the temperature that insures that rearrangements will not occur.
\begin{figure}[!t]
\includegraphics[width=\columnwidth]{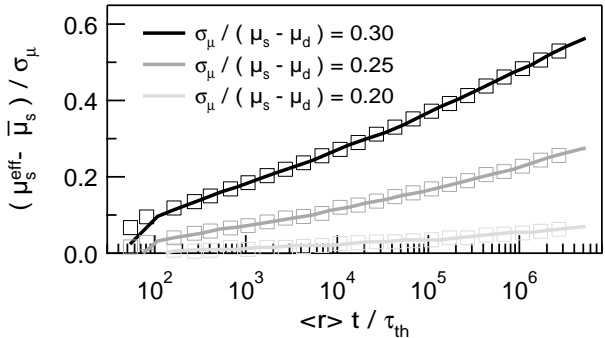}
\caption{Variation $( \mu_{\mathrm{s}}^\mathrm{eff} - \overline{\mu}_{\mathrm{s}})/ \sigma_\mu$ vs.~number ${\langle r\rangle t}/{\tau_\mathrm{th}}$ - We report data obtained for two different values of the dynamical frictional coefficient (Lines : $\mu_\mathrm{d} = 0.5$; Open squares : $\mu_\mathrm{d} = 0.4$). The values of $\sigma_\mathrm{\mu}$ are chosen in accordance with
the values of ${\sigma_\mu}/{(\overline{\mu}_{\mathrm{s}} - {\mu}_{\mathrm{d}})}$ given in the figure ($\overline{\mu}_\mathrm{s} = 0.6$, $\tilde{N} = 10^{-7}$, $\sigma_\theta = 10^{-6}$).}
\label{mueff}
\end{figure}

Experimentally, one possible method to highlight the aging process consists in measuring the critical angle of stability:
the substrate is slowly tilted until the system starts sliding as a whole for a critical tilt angle, $\alpha_c$ \cite{Crassous99}.
In this configuration, the sliders are subjected to normal and tangential forces due to gravity and one can relate $\alpha_c$
with an effective static frictional coefficient $\mu_{\mathrm{s}}^\mathrm{eff} \equiv \tan{\alpha_c}$.
We assessed numerically the temporal evolution of $\mu_{\mathrm{s}}^\mathrm{eff}$ as described in the Appendix \ref{annexe}.
We observe that the increase in $\mu_\mathrm{max}$ (Fig.~\ref{aging}) results in an almost logarithmic increase
in $\mu_{\mathrm{s}}^\mathrm{eff}$ with time $t$.
As suggested by Eq.~(\ref{mumaxth}), we report the relative variation
$(\mu_{\mathrm{s}}^\mathrm{eff} - \overline{\mu}_{\mathrm{s}})/\sigma_\mu$ as a function of the
typical number of rearrangements $n \approx \langle r\rangle t/\tau_\mathrm{th}$ (Fig.~\ref{mueff}).
We observe a collapse of the curves corresponding to the same value of 
$\sigma_\mu/(\overline{\mu}_{\mathrm{s}} - {\mu}_{\mathrm{d}})$, which suggests:
\begin{equation}
\mu_{\mathrm{s}}^\mathrm{eff}(t) = \overline{\mu}_{\mathrm{s}} + \sigma_\mu \Phi\Bigl(\frac{\sigma_\mu}{\overline{\mu}_{\mathrm{s}} - {\mu}_{\mathrm{d}}},\frac{\langle r\rangle t}{\tau_\mathrm{th}}\Bigr)\,.
\label{final}
\end{equation}
From the numerical data, the function $\Phi$ depends almost logarithmically on $t$ at long times.
Asymptotically, as a function of the number of rearrangements $n(t)$, $\mu_{\mathrm{s}}^\mathrm{eff} \sim \overline{\mu}_{\mathrm{s}}
+ \phi\,\sigma_\mu\,(\log n)^\beta$ with $\beta$ of the order of unity and $\phi$ a constant
which only depends on ${\sigma_\mu}/{(\overline{\mu}_{\mathrm{s}} - {\mu}_{\mathrm{d}})}$.
The aging phenomenon is drastically enhanced when the width $\sigma_\mu$ is increased. 

It is particularly interesting to estimate the expected order of magnitude of the effect.
With $\tilde{N} \sim 10^{-8}$ and typical values $\overline{\mu}_{\mathrm{s}} \sim 0.6$ and ${\mu}_{\mathrm{d}} \sim 0.5$ of the frictional coefficients, temperature changes of about 0.01 K correspond to $\sigma_\theta \sim 10^{-7}$ and, thus, to typically $<r> \sim 50$ rearrangements during the characteristic time $\tau_\mathrm{th}$.
From a typical thermal conductivity $\lambda$ of about 1 W/(m K) and a typical heat capacity $C$ of about 10$^6$ J/m$^3$, one can estimate from the diffusion coefficient $\lambda/C$ that $\tau_\mathrm{th} \sim l_0^2 C / \lambda$ ranges from 10$^{-2}$ to 1 s (indeed much larger than $\tau_\mathrm{dyn}$) for a typical size of the system ranging from 1 to 10 cm.
Considering further a value $\sigma_\mathrm{\mu} \sim 0.02$ (about 3 \% percent of the average value
$\overline{\mu}_{\mathrm{s}}$), we expect from Fig.~\ref{mueff} an increase of about 0.2~\% of the effective frictional coefficient after $10^5\,\tau_\mathrm{th} \sim 10^3 - 10^5$~s.
Note however that for $\sigma_\mathrm{\mu} \sim 0.03$ (about 5 \% percent of the average value $\overline{\mu}_{\mathrm{s}}$), the effect is drastically enhanced leading to a change of about
5~\% after the same waiting time.
Thus, small temperature variations are likely to induce, in a few hours, an increase of the effective frictional coefficient of about a few percents.
Note that, from Eq.~(\ref{final}), we expect the aging process to mainly depend on $\sigma_\mathrm{\mu}$ and $\sigma_\mathrm{\theta}$ (through $\langle r \rangle$),
the effect being expected to be larger for broader distribution of the frictional coefficient and for larger amplitude of the temperature variations.
Finally, it is important to notice that the thermal dilations are important because of the change $k l_0 \theta$ they induce in the force between the two sliders and that the latter change is to be compared with the normal force $N$. Thus, a smaller $k$ leads to a weaker effect of the temperature changes. This is probably the reason why no or little aging is observed with soft materials \cite{Crassous99}.

\section{Conclusion}
\label{conclusion}

Reducing the study of the frictional contact between two solid surfaces to the study of two frictional sliders, connected by a spring, in contact with a flat substrate, we demonstrated that the uncontrolled thermal dilations of the system can lead to a significant increase in the effective static frictional coefficient $\mu_{\mathrm{s}}^\mathrm{eff}$ with time.
The evolution of $\mu_{\mathrm{max}}$ at long times is assessed theoretically and the associated $\mu_{\mathrm{s}}^\mathrm{eff}$ is estimated numerically.
Our simplistic model makes it possible to demonstrate that the process leads to an almost logarithmic increase in $\mu_{\mathrm{s}}^\mathrm{eff}$ and we account for the dependence on the parameters, in particular the distribution of the static frictional coefficient associated with the solid-solid contact and the characteristics of the temperature changes (amplitude and characteristic time).
Considering typical orders of magnitude of the physical ingredients, we show that temperature changes can have significant effects, which can also contribute to the aging of the frictional coefficient observed experimentally \cite{Crassous99}.

Finally, one could wonder if the effect reported here would not be specific to the system we chose.
{We assumed that the contact between two flat and rough surfaces reduces to a small number of coherent regions. On the one hand, one could claim that the contact between the surfaces rather consists in a large number of microscopic contact points and that Non-Amonton behavior of the microscopic contacts was revealed experimentally \cite{Bureau06}.}
However, whatever the characteristics of the microscopic contacts (characterized by frictional mechanical properties or not), one can guess that the thermal dilations, which induce a change in the distribution of the force network between the latter, lead to rearrangements of the less stable ones. The process again favors the most stable contacts and we expect the effective frictional coefficient to increase with time.
{On the other hand, even considering coherent regions exhibiting frictional properties, one could
claim that their number is likely to be significantly larger than considered in the present study.
Again, we do not expect an increase in the number
of contact regions to change qualitatively the aging process. However, the quantitative effect is difficult to predict theoretically.  The study of a 2D irregular network of sliders is thus pertinent and will be the subject of a forthcoming publication in which the aging, the diffusion, as well as the creep motion induced by a constant tangential force, will be considered.}  

\appendix
\section{Effective static frictional coefficient}
\label{annexe}

In this appendix, we present the numerical method used to compute the time evolution of the effective frictional coefficient $\mu_{\mathrm{s}}^\mathrm{eff}$. To this aim, we consider that the substrate on which the two sliders lie can be tilted by an ajustable angle $\alpha$.

The dynamics of the system on the horizontal substrate, as described in Sec.~\ref{changes}, is divided in time steps of duration $\tau_\mathrm{th}$. Within each time step, the sliders evolve through a sequence of rearrangements during which the positions and static frictional coefficients change. Before each rearrangement, the values of $\mu_{\mathrm{s},1}$, $\mu_{\mathrm{s},2}$ and of the force
$F_{1\to2}$ are copied to auxiliary variables, and the angle of avalanche corresponding to
this precise set of variables is determined through an iterative algorithm that we now describe.

Due to gravity, when the substrate is tilted, the sliders on the incline are both subjected
to the same positive force $f=mg\sin\alpha$. Correspondingly, the normal force now becomes $N=mg\cos\alpha$.
According to Eq.~(\ref{stability}), the slider 1 (resp. 2) starts moving if one of the two following conditions
are fulfilled:
\begin{subequations}
\label{eq-stab1-app}
\begin{align}
f - F_{1\to2} &> \mu_{\mathrm{s},1} {N},\\
f + F_{1\to2} &> \mu_{\mathrm{s},2} {N}.
\end{align}
\end{subequations}
One deduces from these relations that, when $\alpha$ is increased, the slider 1 moves first if
\be \label{eq-start-motion}
2 F_{1\to2} < (\mu_{\mathrm{s},2}-\mu_{\mathrm{s},1})\,N,
\ee
where $F_{1\to2}$ is given by Eq.~(\ref{eqF12}).
The onset of motion occurs precisely when $f \equiv mg\sin\alpha$
reaches the value $f_1=\mu_{\mathrm{s},1} {N}+F_{1\to2}$.
If the condition (\ref{eq-start-motion}) is not fulfilled, the slider 2 moves first, when $f$ reaches the value
$f_2= \mu_{\mathrm{s},2} {N}-F_{1\to2}$.

The motion of the slider $i$ induces the motion of the other slider if the condition (\ref{stability}) is satisfied, thus if (after simple algebra):
\begin{equation}
\frac{f}{N} \ge \mu_\mathrm{d} + \frac{1}{2} (-1)^i\,( \mu_{\mathrm{s},1} - \mu_{\mathrm{s},2} ).
\label{stab}
\end{equation}
Then, if $f>\mu_\mathrm{d} N$, the motion of the two sliders is accelerated, and a macroscopic sliding of the whole system
is observed. The corresponding value of $\tan \alpha_\mathrm{c} = f/mg$ is recorded.
In the opposite case, when $f<\mu_\mathrm{d}$, the motion is damped, leading only to a short
displacement of the system (creep). Once the system is again at rest,
new values of $\mu_{\mathrm{s},1}$ and $\mu_{\mathrm{s},2}$ are drawn from the
Gaussian distribution $p(\mu_{\mathrm{s}})$. The value of the force $F_{1\to2}$ also needs to be recomputed,
due to the change of the positions $x_1$ and $x_2$. Since the determination of the new positions would require to
integrate the equations of motion during the damped motion, which is time-consuming, we rather use a simple
approximation. The new force $F_{1\to2}$ is chosen at random, with a uniform probability, between
the values $f-\mu_{\mathrm{s},1}\,N$ and $\mu_{\mathrm{s},2}\,N-f$, where $\mu_{\mathrm{s},1}$ and $\mu_{\mathrm{s},2}$
designate the new values of the static frictional coefficient. We are again led back to Eq.~(\ref{eq-stab1-app}),
and the process is then iterated until a macroscopic sliding is observed.

If, however, Eq.~(\ref{stab}) is not satisfied, the other slider remains at rest when the slider $i$ rearranges,
and the release of the elastic energy loaded in the spring leads to a change
\be
\Delta F_{1\to2} = - 2\,(-1)^i ( \mu_{\mathrm{s},i} - \mu_\mathrm{d}) N
\ee
in the force $F_{1\to2}$ between the sliders --see Eq.~(\ref{displacement}).
The value of $\mu_{\mathrm{s},i}$ needs to be drawn anew from $p(\mu_{\mathrm{s}})$, the other frictional coefficient
remaining unchanged, and the process is here again iterated, leading back to Eq.~(\ref{eq-stab1-app}),
as long as a macroscopic motion does not emerge.

At the end of this iteration process, the critical value $\tan \alpha_\mathrm{c}$ has been obtained, and the evolution
on the horizontal substrate resumes (we recall that the evolution on the tilted substrate is only a 'virtual'
test, that does not influence the 'real' horizontal evolution). Obviously, the obtained value of $\tan \alpha_\mathrm{c}$
depends on the values of \{$F_{1\to2}$, $\mu_{\mathrm{s},1}$, $\mu_{\mathrm{s},2}$, $\theta$\} given at the beginning
of the test, which are all stochastic variables. It is thus necessary to average $\tan\alpha_\mathrm{c}$ over the dynamics.
Since, on the other hand, we wish to obtain the time-dependence of the critical angle to study aging effects,
we proceed through a time scale separation as follows:
$\tan \alpha_\mathrm{c}$ is averaged over all the rearrangements occurring 
in a given time step of duration $\tau_{\mathrm{th}}$, and also over a large number of independent realizations
of the horizontal dynamics, with random initial conditions. Then the time-dependence on a scale larger than
$\tau_{\mathrm{th}}$ remains, and the time-step averages $\langle \ldots \rangle_{t_q}$
are labeled by the corresponding time $t_q=q\tau_{\mathrm{th}}$,
with $q$ integer, yielding the effective static frictional coefficient
$\mu_{\mathrm{s}}^{\mathrm{eff}}(t_q) \equiv \langle \tan \alpha_\mathrm{c}\rangle_{t_q}$.


\begin{thebibliography}{99}

\bibitem{Chen09} K. Chen, A. Harris, J. Draskovic and P. Schiffer, Gran. Matt. {\bf 11}, 237 (2009).

\bibitem{Chen06} K. Chen, J. Cole, C. Conger, J. Draskovic, M. Lohr, K. Klein, T. Scheidemantel and P. Schiffer, Nature {\bf 442}, 257 (2006).

\bibitem{Divoux09} T. Divoux, I. Vassilief, H. Gayvallet and J.-C. G\'eminard, AIP Conference Proceedings (Eds. M. Nakagawa and S. Luding),
{\it 6th International Conference on the Micromechanics of Granular Media} (Golden, CO, 2009).

\bibitem{Divoux08} T. Divoux, H. Gayvallet and J.-C. G\'eminard, Phys. Rev. Lett. {\bf 101} 148303 (2008).

\bibitem{Bocquet98} L. Bocquet, E. Charlaix, S. Ciliberto and J. Crassous, Nature {\bf 396}, 735 (1998).

\bibitem{Fraisse99} N. Fraysse, H. Thom\'e and L. Petit, Eur. Phys. J. B {\bf 11}, 615 (1999).


\bibitem{Geminard01} J.-C. G\'eminard, W. Losert and J. P. Gollub, Conference Proceedings (Ed. Y. Kishino) {\it 4th International Conference on the Micromechanics of Granular Media} (Sendai, Japan, 2001).

\bibitem{Losert00} W. Losert, J.-C. G\'eminard and J. P. Gollub, Phys. Rev. E {\bf 61}, 4060 (2000).

\bibitem{Gayvallet02} H. Gayvallet and J.-C. G\'eminard, Eur. Phys. J. B {\bf 30}, 369 (2002). 

\bibitem{Bowden50} F. P. Bowden and D. Tabor, {\it Friction and Lubrication of Solids}
(Clarendon, Oxford, 1950).

\bibitem{Rabinowicz65} E. Rabinowicz, {\it Friction and Wear of Material}
(Wiley, New York, 1965).

\bibitem{Rabinowicz58} E. Rabinowicz, Proc. Phys. Soc. London {\bf 71}, 668 (1958).

\bibitem{Scholz90} C. H. Scholz, {\it The Mechanics of Earthquakes and Faulting} (Cambridge University, Cambridge, England, 1990).

\bibitem{Heslot94} F. Heslot, T. Baumberger, B. Perrin, B. Caroli and C. Caroli, Phys. Rev. E {\bf 49} 4973 (1994).

\bibitem{Crassous99} 
J. Crassous, L. Bocquet, S. Ciliberto and C. Laroche, Europhys. Lett. {\bf 47}, 562-567 (1999).

\bibitem{Berthoud99} P. Berthoud, T. Baumberger, C. G'Sell and J.-M. Hiver, Phys. Rev. B {\bf 59}, 14 313 (1999).

\bibitem{Bureau02} L. Bureau, T. Baumberger and C. Caroli, Eur. Phys. J. E {\bf 8}, 331 (2002).

\bibitem{Tomlinson29} G. A. Tomlinson, Phil. Mag. {\bf 7} 905 (1929).

\bibitem{Joanny84} J.-F. Joanny and P.-G. de Gennes, J. Chem. Phys. {\bf 11}, 552 (1984).

\bibitem{Caroli96} Caroli C. and Nozi\`eres P., Vol. 311 of NATO Advanced Study
Institute, series E (Eds. B.N.J. Persson and E. Tosatti),  {\it Physics of sliding surfaces} (Kluwer Academic publishers, Dordrecht, 1996).

\bibitem{Crassous97} J. Crassous, S. Ciliberto, E. Charlaix and C. Laroche, J. Phys. II France {\bf 7}, 1745 (1997).

\bibitem{Olofsson95} U. Olofsson, Tribology Int. {\bf 28}, 207 (1995).

\bibitem{Persson10} B. N. J. Persson, J. Phys.: Condens. Matter {\bf 22},  265004 (2010).

\bibitem{Baumberger06} T. Baumberger and C. Caroli,
Adv. Phys. {\bf 55}, 279-348 (2006). 

\bibitem{BarratMezard}
A. Barrat and M. M\'ezard, J. Phys. I (France) {\bf 5}, 941 (1995).

\bibitem{Bertin10}
E. Bertin, J. Phys. A: Math. Theor. {\bf 43}, 345002 (2010).

\bibitem{Gumbel}
E.~J. Gumbel, {\it Statistics of Extremes} (New York, Columbia University
Press, 1958; Dover publication, 2004).

\bibitem{Galambos}
J. Galambos, {\it The Asymptotic Theory of Extreme Order Statistics}
(New York, Wiley, 1987).

\bibitem{Krug}
J. Krug, J. Stat. Mech. (2007) P07001.

\bibitem{Bureau06} L. Bureau, T. Baumberger and C. Caroli, Eur. Phys. J. E {\bf 19}, 163 (2006).

\end{thebibliography}
\end{document}